\documentclass[12pt]{article}

\input epsf
\usepackage{amssymb}
\usepackage[dvips]{graphicx}

\def\npb#1#2#3{{\it Nucl. Phys.} {\bf B#1} (#2) #3 }
\def\plb#1#2#3{{\it Phys. Lett.} {\bf B#1} (#2) #3 }
\def\prd#1#2#3{{\it Phys. Rev. } {\bf D#1} (#2) #3 }
\def\prl#1#2#3{{\it Phys. Rev. Lett.} {\bf #1} (#2) #3 }
\def\mpla#1#2#3{{\it Mod. Phys. Lett.} {\bf A#1} (#2) #3 }

\def\bb#1{{\tt hep-th/#1}}
\def\grqc#1{{\tt gr-qc/#1}}
\def\heph#1{{\tt hep-ph/#1}}

\def\rmp#1#2#3{{\it Rev. Mod. Phys.} {\bf #1} (#2) #3 }

\def\cqg#1#2#3{{\it Class. Quantum Grav. } {\bf #1} (#2) #3 }
\def\apj#1#2#3{{\it Astrophys. J.  } {\bf #1} (#2) #3 }

\def\jhep#1#2#3{{\it J. High Energy Phys.} {\bf #1} (#2) #3 }

\def\aph#1{{\tt astro-ph/#1}}

\def\nat#1#2#3{{\it Nature} {\bf #1} (#2) #3 }
\def\ijtp#1#2#3{{\it Int. J. Theor. Phys.} {\bf #1} (#2) #3 }
\def\prep#1#2#3{{\it Phys. Rep.} {\bf #1} (#2) #3 }
\def\aj#1#2#3{{\it Astron. J.} {\bf #1} (#2) #3 }

   \def\CG{{\cal G}}
   
 \def\CR{{\cal R}}  
   \def\CV{{\cal V}}

\def\dj{\hbox{d\kern-0.347em \vrule width 0.3em height 1.252ex depth
-1.21ex \kern 0.051em}}

\def\half{{1\over 2}\,}

\begin{document}

\setlength{\oddsidemargin}{0cm}
\setlength{\baselineskip}{7mm}


\thispagestyle{empty}
\setcounter{page}{0}

\begin{flushright}
{\tt hep-th/0109038}
\end{flushright}

\vspace*{2cm}

\centerline{\huge \bf Quintessential brane cosmology}

\vskip 2cm

\centerline{K.E. Kunze\footnote{E-mail: kunze@amorgos.unige.ch} }

\vskip 0.3cm

\centerline{{\sl 
D\'epartement de Physique Th\'eorique, 
Universit\'e de Gen\`eve}} 
\centerline{{\sl  24 Quai Ernest Ansermet, CH-1211 Gen\`eve 4, Switzerland}}

\vskip 0.5cm

\centerline{{and}}

\vskip 0.5cm

\centerline{M.A. V\'azquez-Mozo\footnote{E-mail: vazquez@nbi.dk.}$^{,}$\footnote{Address after January 1st, 2002: Theory
Division, CERN, CH-1211 Geneva 23, Switzerland.}}

\vskip 0.3cm

\centerline{{\sl Niels Bohr Institute, Blegdamsvej 17,}}
\centerline{{\sl DK-2100 Copenhagen \O, Denmark}}

\vskip 1cm

\noindent  
We study a class of braneworlds where the cosmological evolution arises as the result of
the movement of a three-brane in a five-dimensional static dilatonic bulk, with and
without reflection symmetry. The resulting four-dimensional Friedmann equation 
includes a term which, for a certain range of the parameters, effectively works as a quintessence 
component, producing an acceleration of the universe at late times. 
Using current observations and bounds derived from big-bang nucleosynthesis we 
estimate the parameters that characterize the model.

\vskip 1cm

\noindent
September 2001

\newpage

\section{Introduction and motivation}
\setcounter{equation}{0}

One of the most interesting experimental results in cosmology in the last years has 
been the measure of the acceleration of the 
universe from the observation of high-redshift supernovae \cite{perl}. 
The combination of this result with the
measure of the cosmic microwave background (CMB) anisotropies \cite{cmb} favors a universe whose energy density is
very close to the critical one, 
$\Omega_{\rm total}\simeq 1$, and with a large 
dark energy component, $\Omega_{\rm dark}\simeq 0.7$. 
This realization that the energy content of the universe might be dominated by some 
kind of dark energy has revived the interest in cosmological models where the late time
expansion  of the universe is governed by some kind
of vacuum energy, either in the form of a  
nonvanishing cosmological constant or a quintessence field (see \cite{cc,Qbine} 
for comprehensive reviews). 

Recently, braneworld scenarios have emerged as a  very appealing 
alternative to standard dimensional reduction in higher-dimensional theories. 
The general idea is the following. The observable universe is a submanifold 
embedded in a higher-dimensional space-time.
By some physical mechanism matter fields are confined to this submanifold.
Although gravity can propagate into the bulk, the Kaluza-Klein
excitations of the four-dimensional graviton can be decoupled 
at low energies either by considering 
finite but ``large" extra dimensions \cite{dimop} 
or a negative cosmological constant in the 
bulk \cite{rs}. Therefore, at energies below the mass gap of 
these Kaluza-Klein modes,
physics on the brane is well described by the matter degrees of freedom 
confined to the brane and standard gravity in four dimensions. One of the most interesting
features of these models is that the energy scale at which the five-dimensional effects are
important can be relatively low, even of the order of the TeV, and therefore 
in principle accessible to future high-energy experiments.

It has been realized that the braneworld paradigm in its cosmological
application leads to
corrections to the four-dimensional Einstein equations on the brane \cite{sms,mw,mb}. These
departures from General Relativity, and their imprints in braneworld cosmological models,
might be used to test the scenario against cosmological observations. In the context of
Friedmann-Robertson-Walker cosmologies, the braneworld hypothesis results in the
presence of two extra terms 
on the right-hand side of the four-dimensional
Friedmann equation: a dark-radiation contribution and a quadratic term in the energy density
that couples directly to the five-dimensional Planck scale. While the first one might
have any effect at the present stage of the history of the universe, the second one
is expected to be important only in the very early universe where the energy density was
very large.

Quintessence models have also been studied in the context of brane cosmologies 
\cite{bcqui,mizmae}. With a few exceptions \cite{exc}, in most of the models considered 
a four-dimensional decaying cosmological constant is implemented by 
introducing an additional scalar field in four dimensions. 
By a suitable choice of 
potential its contribution 
to the energy density becomes dominant only at present.

In this paper, we  study the possibility within a braneworld
scenario that an effective
four-dimensional quintessence is induced 
by a dilaton field propagating in the bulk. 
In our model the cosmological evolution
in four dimensions reflects the movement of the braneworld in a static 
dilatonic bulk space-time in which $\mathbb{Z}_{2}$-reflection 
symmetry around the location of the brane can be broken. 
We find  that the bulk dilaton induces
a term in the effective four-dimensional Einstein equations that acts
as a decaying cosmological constant. Hence, in this sense
quintessence occurs naturally in this class of models. 
For universes filled with matter and radiation, and for 
a certain range of the parameters defining the model, 
this vacuum energy actually tracks  the matter or radiation energy density
and dominates at late times, 
driving an accelerating expansion 
of the universe. 
Using constraints from big-bang nucleosynthesis and
the present value of Newton's constant allows us to constrain the
parameters of our class of models. In particular, we find 
a lower bound of $10^5\, {\rm GeV}$ for the five-dimensional
Planck mass.  

In the solutions showing quintessencelike behavior 
the universe first undergoes decelerated expansion and
then at late times enters into a regime of accelerated
expansion due to the effects of the bulk dilaton.
In this case, the observed isotropy of the universe has to be assumed.
However, as will be seen, the class of models under study
also allows for inflationary solutions at early times. 
In this type of solution the universe starts in a state
of accelerated expansion and then later undergoes
decelerated expansion. The problem in this kind of models, however, 
is not only the lack of acceleration in the expansion at late times, but the fact
that in this asymptotic regime the decelerated expansion is dominated by the $\rho^2$ term
in the Friedmann equation. 

Much of the recent interest in brane cosmologies was triggered by Ho\v{r}ava-Witten type models
\cite{hw} where the brane is confined 
to lie on the fixed points of an $\mathbb{S}^{1}/\mathbb{Z}_{2}$
orbifold. Thus usually a $\mathbb{Z}_{2}$ symmetry in the 
bulk direction is assumed. However here we will consider more general
situations in which the bulk space-time is not symmetric under reflection 
around the position of the brane. In our context, this situation describes braneworlds
that emerge as domain walls separating two regions of space-time
with different values of the cosmological constant. As we will see, this asymmetry 
leaves its imprints in the cosmological dynamics on the brane at late times, for example
through corrections to the four-dimensional Newton's constant depending on 
the difference between the cosmological constants on the two sides of the brane.

The plan of the paper is as follows. 
In Section 2 we study the dynamics of moving branewolds in 
the background of five-dimensional Einstein-Liouville solutions. 
Section 3 will be 
devoted to the study of the cosmological dynamics of moving brane 
universes with ${\mathbb{Z}}_{2}$-reflection
symmetry, whereas in Section 4 we will analyze brane models where 
this reflection symmetry is broken. In Section 5 we will proceed to study the 
phenomenology of these models by using both current observational data and 
bounds derived from big-bang nucleosynthesis. 
Finally, in Section 6, we will summarize our results.

\section{Moving branes in a static bulk geometry}
\setcounter{equation}{0}

In Ref. \cite{fkvmbr} a broad class of braneworld cosmologies coupled to a dilaton field in the bulk was
studied by using a solution generating algorithm that enables the construction of five-dimensional
Einstein-Liouville cosmologies (cf. \cite{jim}). Assuming ${\mathbb{Z}}_{2}$-reflection symmetry around the 
fixed location of the brane, it was found that the solutions studied enjoy a new version of 
the self-tuning mechanism of  
\cite{st}, in which the total cosmological constant in four-dimensions is canceled independently of the 
value of the bulk cosmological constant and the energy density on the brane.
It was also found there that inflation in Friedmann-Robertson-Walker and Bianchi-I 
models belonging to the family of solutions studied cannot be achieved for any value of the 
amplitude and slope of the bulk Liouville potential. 

Here we are going to use these bulk solutions in a different physical setup.
Our aim is the construction of realistic four-dimensional brane cosmologies
in which the cosmological expansion accelerates at late times. For this, even if
we restrict the bulk solutions to those studied in \cite{fkvmbr}, we will have
to relax some of the conditions imposed on the braneworlds constructed there. 
In particular, here we will not consider the braneworld to have a fixed location on the
bulk, but we will take into account the degree of freedom associated with its collective 
motion in the five-dimensional space-time. In addition, we will also 
consider situations in which $\mathbb{Z}_{2}$-reflection symmetry is broken
(cf. \cite{brz2,derdol,davis}). As we will see in the following, the features of
the resulting brane cosmologies are quite different from the ones studied in \cite{fkvmbr}.

\subsection{Description of the model}

We will consider solutions to the five-dimensional bulk action 
\begin{eqnarray}
S_{\rm 5D}&=&{1\over \kappa_{5}^2}\int d^{5}x\sqrt{-\CG}\left[{1\over 2}\CR -\half (\partial\phi)^{2}
-\Lambda e^{-{2\over 3}k\phi} \right] \nonumber \\
&+&\int_{Y} d^{4}x\sqrt{-g}\left[{1\over \kappa_{5}^{2}}K^{\pm}
-\lambda(\phi)+e^{4b\phi}L(e^{2b\phi}g_{\mu\nu})_{\rm matter}\right].
\label{act}
\end{eqnarray}
Here we are assuming that the five-dimensional Planck scale $\kappa_{5}$ 
determines also the scale
of the five-dimensional scalar field, which is dimensionless here. The only other independent 
energy scales 
entering into the action (\ref{act}) are
provided by $\Lambda$, which has dimensions of $({\rm mass})^2$, and the energy scale 
of the four-dimensional
dynamics. In the following, we will consider that the matter on the brane is described by 
a perfect fluid with 
barotropic equation of state $p=w\rho$ (with $|w|\leq 1$), 
so the matter Lagrangian is written in terms of the energy-density as
$L_{\rm matter}=-\rho$. In \cite{fkvmbr}, solutions to the equations of motion in the bulk 
derived from (\ref{act}) were
constructed in terms of the functions $Q(x)$ and $h_{\mu\nu}(x)$ in which the metric takes 
the form\footnote{These
solutions are related to the ones studied in \cite{fkvmbr} by a $k$-dependent 
coordinate redefinition
$r^2=\exp{(a_{2}\chi)}$ and a shift in the dilaton field.}.
\begin{equation}
ds^{2}_{\rm 5D} = {1\over \xi^2}e^{{4k\over \sqrt{k^2+6}}Q(x)}\,r^{{2\over 3}(k^2-3)}\,dr^2+ 
e^{-{2k\over \sqrt{k^2+6}}Q(x)}\,r^{2}\,h_{\mu\nu}(x)\,
dx^{\mu}dx^{\nu} 
\label{lefd}
\end{equation}
with the scalar field given by
\begin{equation}
\phi(x,r)={6\over \sqrt{k^2+6}}\,Q(x)+k\log{r}.
\label{dil}
\end{equation}
The metric function $h_{\mu\nu}$ is a solution to the four-dimensional Einstein equations coupled
to a scalar field, $R_{\mu\nu}=\partial_{\mu}\psi\partial_{\nu}\psi$ with $\psi(x)=\sqrt{6}Q(x)$.
The amplitude of the Liouville potential as a function of $k$ and $\xi$ has the following form 
\begin{equation}
\Lambda = {1\over 2}(k^2-12)\xi^2.
\end{equation}
The parameter $\xi$ has dimensions of mass and determines the scale of the cosmological constant.

Here we will study the motion of a braneworld in a bulk geometry described by the 
line element (\ref{lefd}), possibly with different values of $\xi$ on the two sides of the brane. 
In order to simplify our analysis, we consider $Q(x)\equiv 0$
in Eq. (\ref{dil}) and the bulk space-time to be static. This implies in particular that $h_{\mu\nu}$ is
a static Ricci-flat four-dimensional metric. 
Hence the metric takes the form
\begin{equation}
ds_{\rm 5D}^2={1\over \xi^2}
r^{{2\over 3}\left(k^2-3\right)}dr^2+r^2\,h_{\mu\nu}dx^{\mu}dx^{\nu},
\label{bR}
\end{equation}
while the dilaton is  
\begin{equation}
\phi=k\log r.
\label{dilR}
\end{equation}
In addition, the four-dimensional metric is given by
\begin{equation}
h_{\mu\nu}dx^{\mu}dx^{\nu}=-dt^2+h_{ij}(x)dx^{i}dx^{j}.
\label{metbr}
\end{equation}

Let us then consider a single braneworld embedded in the five-dimensional space-time
described by Eq. (\ref{bR}) and moving along the trajectory
\begin{equation}
t=t(\tau),\hspace*{1cm} r=R(\tau),\hspace*{1cm}x,y,z={\rm constant},
\end{equation}
where $\tau$ is the proper time. For a five-dimensional static metric of the form 
$ds^2=\CG_{55}dr^2+\CG_{00}dt^2+\CG_{ij}dx^{i}dx^{j}$, 
the 
nonvanishing components of the 5-velocity $\CV^{A}$ and the vector normal to the
brane $n^{A}$ can be computed using the relations
$\CV_A \CV^A=-1$, $\CV_A n^A=0$, with the result
\begin{eqnarray}
\CV^0 &=& \sqrt{1+\CG_{55}\dot{R}^2\over -\CG_{00}}, \hskip 1.3cm
\CV^{5}=\dot{R}\nonumber \\
n^0 &=& \dot{R}\sqrt{\CG_{55}\over -\CG_{00}}, \hskip 2cm
n^{5}=\sqrt{1+\CG_{55}\dot{R}^2\over \CG_{55}}.
\label{e}
\end{eqnarray}
Here and in the rest of the paper
the dots indicate differentiation with respect to proper time $\tau$.

On the other hand, taking an orthonormal frame $\{\CV_M, n_M, e^{(i)}_{\,\,\,M}\}$, the components of the
extrinsic curvature can be easily computed using the existence of the timelike Killing vector of 
the static bulk space-time (cf. \cite{visL,charea}):
\begin{eqnarray}
K_{\tau}^{\tau}&=&{1\over \dot{R}\sqrt{-\CG_{00}\CG_{55}}}
{d\over d\tau}\left[\sqrt{-\CG_{00}}\sqrt{1+\CG_{55}\dot{R}^2}\,
\right],\\
K^{i}_{j}&=&{1\over 2}n^{M}\CG^{ik}\partial_M \CG_{kj}.
\label{et}
\end{eqnarray}
Note that $K_{\tau\tau}$ is given by the orthogonal component
of the proper acceleration of the
brane. 

Looking at Eqs. (\ref{bR}) and (\ref{metbr}), we see that, when expressed in proper time, the induced metric 
on the brane takes the form
\begin{equation}
ds^2_{\rm 4D}=-d\tau^2+R^{2}(\tau)h_{ij}(x)dx^i dx^j.
\label{metind}
\end{equation}
Since the metric (\ref{metbr}) has to be a static vacuum solution to
Einstein equations, in the following we will focus on the simplest possibility where 
$h_{ij}=\delta_{ij}$. In this case the line element (\ref{metind}) describes a flat 
Friedmann-Robertson-Walker universe.

\subsection{Junction conditions and dynamical equations}

As explained above, we will consider that 
the constant $\xi$ determining the amplitude of the Liouville 
potential takes in
general different values 
$\xi_{\pm}$ on the two sides of the brane (a related proposal has been studied
in  \cite{derdol}). This means that the bulk geometry is described by the line elements 
\begin{eqnarray}
ds^{2}_{\rm 5D}\Big|_{+}&=&{1\over\xi_{+}^2}r^{{2\over 3}(k^2-3)}\,dr^2+r^{2}[-dt^2+\delta_{ij}dx^i dx^{j}],
\hskip 2cm r>R(\tau), \nonumber \\
ds^{2}_{\rm 5D}\Big|_{-}&=&{1\over\xi_{-}^2}r^{{2\over 3}(k^2-3)}\,dr^2+r^{2}[-N(t)^2\,dt^2+\delta_{ij}dx^i dx^{j}],
\hskip 1.1cm r<R(\tau).
\end{eqnarray}
Following \cite{derdol} we have introduced a shift function $N(t)$ in order to ensure the continuity of the 
metric across the brane. Notice that the bulk field equations are satisfied since, given the static character of 
the metric, the shift function can be locally reabsorbed in a redefinition of time. A simple calculation shows that 
the metric is continuous across $r=R(\tau)$ by taking 
\begin{equation}
N(t)^2=\left({\xi_{-}\over \xi_{+}}\right)^2{\xi_{+}^2+R^{{2\over 3}(k^2-3)}\dot{R}^{2}\over
\xi_{-}^2+R^{{2\over 3}(k^2-3)}\dot{R}^{2}}\,.
\end{equation}
On the other hand, the dilaton is continuous across the location of the brane.

The Israel junction conditions, together with the matching for the
dilaton across the brane, will determine the dynamics of 
the scale factor $R(\tau)$. As for the metric junction equations we find
\begin{equation}
[K_{\mu\nu}]\equiv K^{+}_{\mu\nu}-K^{-}_{\mu\nu}
={\kappa_{5}^2\over 3}\left[-\lambda(\phi)+g^{\alpha\beta}\tau_{\alpha\beta}\right]\,g_{\mu\nu}-\kappa_{5}^2 \tau_{\mu\nu},
\end{equation}
where $g_{\mu\nu}$ is the induced metric, $\tau_{\mu\nu}$ is the energy-momentum tensor on the brane 
and $\lambda(\phi)$ is the brane 
tension. Indices are raised and lowered 
using $g_{\mu\nu}$. Since we are assuming that the matter on the brane is a perfect fluid the 
energy-momentum tensor is given by $\tau^{\mu}_{\nu}={\rm diag}(-\rho,p,p,p)$. Using the 
barotropic equation of state $p=w\rho$ we find from the matching condition for $K^{\tau}_{\tau}$,
\begin{eqnarray}
& &\hspace*{-1cm}\left[\left({\dot{R}\over R}\right)^2+
\xi_{+}^2 R^{-{2\over 3}k^2}
\right]^{-{1\over2}}
\left[{\ddot{R}\over R}+
{k^2\over 3}\left({\dot{R}\over R}\right)^2
+\xi_{+}^2 R^{-{2\over 3}k^2}
\right] \label{eh}\\
&-&\left[
\left({\dot{R}\over R}\right)^2+
\xi_{-}^2 R^{-{2 \over 3}k^2}
\right]^{-{1\over2}}
\left[{\ddot{R}\over R}+
{k^2\over 3}\left({\dot{R}\over R}\right)^2
+\xi_{-}^2 R^{-{2 \over 3}k^2}
\right]=-{\kappa_{5}^2\over 3}{\lambda(\phi)}+\kappa_{5}^2\left(w+{2\over 3}\right)\rho
\nonumber 
\end{eqnarray}
while from the jump of $K^{i}_{j}$ at $r=R(\tau)$ we get
\begin{equation}
\left[
\left({\dot{R}\over R}\right)^2+
\xi_{+}^2 R^{-{2\over 3}k^2}
\right]^{1\over2}
-\left[
\left({\dot{R}\over R}\right)^2+
\xi_{-}^2 R^{-{2\over 3}k^2}
\right]^{1\over2}
=-{\kappa_{5}^2\over3}[\lambda(\phi)+\rho].
\label{ehh}
\end{equation}
In addition to this, we have also to impose  
the matching condition for the dilaton derivative across the 
brane \cite{charea},
\begin{equation}
\left[n^{A}\partial_{A}\phi\right]=\kappa_{5}^2\left[\lambda'(\phi)-b\,g^{\mu\nu}\tau_{\mu\nu}\right]
\end{equation}
which, upon substitution, reads
\begin{equation}
k\left[
\left({\dot{R}\over R}\right)^2+
\xi_{+}^2 R^{-{2\over 3}k^2}
\right]^{1\over2}
-k\left[
\left({\dot{R}\over R}\right)^2+
\xi_{-}^2 R^{-{2\over 3}k^2}
\right]^{1\over2}
={\kappa_{5}^2}\left[\lambda'(\phi)-(3w-1)b\,\rho\right].
\label{em}
\end{equation}

The particular case in which the bulk space-time has ${\mathbb{Z}}_2$-reflection 
symmetry across the position of the brane is recovered by 
setting $\xi_{+}=\xi_{-}\equiv{\xi}$ and taking the outward normals 
on the two sides of the brane with opposite signs. In Eqs. (\ref{eh})-(\ref{em}), this amounts to
replacing the minus sign between the two terms on the left hand side of these equations by a 
plus sign. 

Equations (\ref{eh})-(\ref{em}) derived from the junction conditions determine the dynamics of the
four-dimensional scale factor $R(\tau)$ in terms of the brane tension $\lambda(\phi)$ and the
energy density $\rho$. Consistency of Eqs. (\ref{ehh}) and (\ref{em}) implies the following 
differential equation relating the brane tension with the energy density:
\begin{equation}
\lambda'(\phi)+{k\over 3}\lambda(\phi)=\left[(3w-1)\,b-{k\over 3}\right]\,\rho.
\label{dffeq}
\end{equation}
It is remarkable that, because of the junction condition for the dilaton, the matter energy 
density on the brane actually mixes with the brane tension. 

It is important to stress that any solution of Eq. (\ref{ehh}) will automatically satisfy the second
order differential equation derived from Eq. (\ref{eh}), provided the brane tension $\lambda(\phi)$ and
the density $\rho$ satisfy (\ref{dffeq}) and (\ref{ce}). Thus, as it is the case in standard cosmology, 
the dynamics of the brane universe is solely determined by the first-order differential
equation (\ref{ehh}).

The analysis of the case $k=0$ requires some care. Since for vanishing $k$ the dilaton field 
exactly vanishes for the family of solutions studied, the dilaton matching equation (\ref{em}) is
not imposed and the only dynamical equation is Eq. (\ref{ehh}) with constant brane tension 
$\lambda(\phi)=\lambda_{0}$. In this case, therefore, there is no mixing between the 
vacuum energy and the matter on the brane.

\subsection{Energy balance on the brane}

In order to get the form of the brane tension $\lambda(\phi)$ from Eq. (\ref{dffeq}) it would 
be necessary to know how the energy density depends on the scale factor 
(or equivalently, for $k\neq 0$,
on the dilaton field).
The conservation equations for the energy-momentum on the brane can be 
derived from the Codazzi equation. Actually, as explained in \cite{mw}, in the presence of a dilaton field there is an 
interchange
of energy-momentum between the brane and the bulk. This implies the existence of
a nonvanishing right-hand side in the conservation equation
\begin{equation}
\dot{\rho}+3{\dot{R}\over R}(\rho+p)
=-(3p-\rho)\,b\,\dot{R}{d\phi\over dR}.
\label{eee}
\end{equation}
We see that the source term in the continuity 
equation results from the conformal coupling between the matter-energy density on the brane
and the bulk dilaton, measured by $b$. 

Using the equation of state of the fluid and the expression of the dilaton field in terms
of the scale factor, Eq. (\ref{dilR}), we find the following differential equation for $\rho(R)$: 
\begin{equation}
R{d\rho\over dR}+m \rho=0, \hskip 1cm {\rm with}\hskip 1cm  
m \equiv 3(w+1)+(3w-1)\,k\,b.
\label{ce}
\end{equation}
This implies that the energy density depends on the scale factor and the dilaton field as
\begin{equation}
\rho(R)=\rho_{0} R^{-m} \hskip 0.5cm \Longrightarrow \hskip 0.5cm 
\rho(\phi)=\rho_{0} \,e^{-{m\over k}\phi},
\label{eerho}
\end{equation}
where the second equation is only valid for $k\neq 0$.
By comparing these expressions with the ones encountered in ordinary cosmology, we find that
the usual evolution of the energy density with the scale factor is corrected by a term that depends
on the conformal coupling $b$ between the dilaton and the matter fields.
Equation (\ref{eerho}) can be physically interpreted by considering that, for an observer on the brane the source term
in Eq. (\ref{ee}) leads to an effective barotropic index
$w_{\rm eff}$ governing the evolution of the energy density with the scale factor 
$\rho(R)\sim R^{-3(w_{\rm eff}+1)}$,
where
\begin{equation}
w_{\rm eff}=w+\left(w-{1\over 3}\right)kb.
\label{meos}
\end{equation}

In Section 5, we will see that for 
phenomenologically viable models $k^2\lesssim 10^{-2}$. Since it is natural to expect 
the coupling $b$ to be of the same order as $k$, the difference between $w$ and 
$w_{\rm eff}$ would be also of order $10^{-2}$.
There are, however, two  particular instances in which the correction to $w$ in Eq. (\ref{meos}) vanishes for any value of $b$.
The first one is when $k=0$, which corresponds to switching the dilaton off, so the right hand side of Eq. (\ref{eee}) is zero
and the usual continuity equation for $\rho(R)$ is recovered. The second one corresponds to the case in which 
we consider a radiation fluid on the brane, $w={1\over 3}$, where again the right-hand side of Eq. (\ref{eee}) vanishes.
This is a consequence of the classical conformal invariance of radiation, which is therefore insensitive to the
conformal coupling to the dilaton. As a result, when the matter on the brane is dominated by a gas of 
massless particles, the temperature evolves with the scale factor in the same way as in 
radiation-dominated standard cosmology, $T(R)\sim R^{-1}$.

By plugging the expression for $\rho(\phi)$ into the right hand side of Eq. (\ref{dffeq}) we find a differential
equation for the brane tension. When $k^2\neq 0, 3m$, the solution is given by
\begin{equation}
\lambda(\phi)=\lambda_0\,e^{-{k\over 3}\phi}-\left[1+{9(w+1)\over k^2-3m}\right]
\,\rho_{0}\,e^{-{m\over k}\phi}.
\label{ee}
\end{equation}
where $\lambda_{0}$ is an integration constant. In the particular case when $k^2=3m$ the solution 
takes the form
\begin{equation}
\lambda(\phi)_{k^2=3m}=\left[\lambda_{0}-{3\over k}(w+1)\rho_{0}\phi\right]e^{-{k\over 3}\phi}.
\label{cp}
\end{equation}
Finally, if $k=0$ the brane tension is constant and does not mix with the matter-energy density,
\begin{equation}
\lambda(\phi)_{k=0}=\lambda_{0}.
\end{equation}
In the following, unless stated otherwise, we will only study the case when $k^2\neq 3m$.

As pointed out above, it is interesting that the expressions (\ref{ee}) and (\ref{cp}) for the
brane tension  depend on the matter energy density through the constant $\rho_{0}$. 
This mixing between matter and vacuum energies on the brane can be seen to be dilaton-mediated,
in the sense that it is produced by the junction conditions of the dilaton on the brane, which couples
to both the brane vacuum energy (through its dependence on $\phi$) and the matter fields in the 
action (\ref{act}). 
Only when $k=0$ is the dilaton switched off and there is no mixing between 
the matter and the vacuum energy on the brane.

\section{Moving braneworlds in a $\mathbb{Z}_{2}$-symmetric bulk space-time}
\setcounter{equation}{0}

In this section, we are going to analyze the cosmological dynamics of a four-dimensional
braneworld moving on a bulk space-time with ${\mathbb{Z}}_{2}$-reflection symmetry around the location
of the brane. This case is of some interest, since it represents a deformation of the 
cosmological models studied in \cite{fkvmbr} by taking into account the 
degrees of freedom associated with the movement of the brane
in the bulk.

The dynamical equations for the scale factor are obtained from Eq. (\ref{ehh}) by taking 
$\xi_{+}=\xi_{-}\equiv \xi$ and reversing the minus sign on the left hand side. This results in 
\begin{equation}
\left({\dot{R}\over R}\right)^2=-\xi^2 R^{-{2\over 3}k^2}+{\kappa_{5}^{4}\over 36}(\lambda+\rho)^2.
\label{pfe}
\end{equation}
As explained in the previous section, there is a mixing between the brane tension 
$\lambda(\phi)$ and the energy density due to the mediating dilaton. This implies that
the expression of the tension as a function of the dilaton field (\ref{ee}) contains a term 
depending on $\rho$. If we collect all the $\rho$-dependent terms in Eq. (\ref{pfe}) together 
we can write the Friedmann-like equation ($k^2\neq 3m$)
\begin{eqnarray}
\left({\dot{R}\over R}\right)^2 & =&{1\over 36}(\kappa_{5}^4\lambda_{0}^2-36\xi^2)R^{-{2\over 3}k^2}
+{\kappa_{5}^{4}\over 18}\left(\pm\lambda_{0}R^{-{k^2\over 3}}\right)\left|{9(w+1)\over k^2-3m}\right|\rho(R) \cr
&+&{\kappa_{5}^{4}\over 36}\left[{9(w+1)\over k^2-3m}\right]^2\rho(R)^2,
\label{mpfe}
\end{eqnarray}
where the $\pm$ signs correspond to $k^2<3m$ and $k^2>3m$, respectively. Let us also remark here 
that in order for the (time-dependent) Newton constant to be positive, we have to restrict 
$\lambda_{0}>0$ when $k^2<3m$ and $\lambda_{0}<0$ if $k^2>3m$. 

When $w>-1$ we can transform Eq. (\ref{mpfe}) into the usual Friedmann equation in brane cosmology 
(cf. \cite{bdl,bdel}) by defining the effective energy density 
in four dimensions through
the rescaling,
\begin{equation}
\rho(R)_{\rm eff}=\left|{9(w+1)\over k^2-3m}\right|\rho(R),
\end{equation}
which amounts to a redefinition of the (positive) integration constant $\rho_{0}$ in Eq. (\ref{eerho}).
In physical terms, this means that the energy density measured by a four-dimensional observer by
looking at the rate of expansion differs from the ``bare'' value that appears in the 
action (\ref{lefd}) by a positive multiplicative constant. This effective energy density satisfies the
same continuity equation as $\rho(R)$ and scales in the same way with the scale factor. 
We will identify it with the energy density driving the cosmological evolution of our 
brane universe.

In order to simplify the notation in the following, we will drop the subscript ``eff'' and denote
the effective four-dimensional energy density by $\rho(R)$. Doing so we finally arrive at the Friedmann equation on the brane,
\begin{equation}
\left({\dot{R}\over R}\right)^2={8\pi \over 3}G_{N}(R)\rho_{Q}(R)+{8\pi\over 3}G_{N}(R)\rho(R)
+{\kappa_{5}^{4}\over 36}\rho(R)^2.
\label{fe}
\end{equation}
We see that, in addition to the usual terms linear and quadratic in $\rho(R)$, there is
also a ``quintessence'' energy density given by
\begin{equation}
\rho_{Q}(R)={\kappa_{5}^4\lambda_{0}^2-36\xi^2 \over 2\kappa_{5}^4 |\lambda_{0}|}R^{-{k^2\over 3}}.
\label{quint}
\end{equation}
As we did in Eq. (\ref{meos}), we can define  the  effective barotropic 
index associated with the quintessence term that determines the scaling of
$\rho_{Q}$ with the scale factor as
\begin{equation}
w_{Q}={k^2\over 9}-1.
\label{wQ}
\end{equation}
Thus, for $k^2\leq 6$, it satisfies $-1\leq w_{Q}\leq -{1\over 3}$ and represents a real 
quintessence contribution to the Friedmann equation.
In addition to this, the four-dimensional effective gravitational coupling in our braneworlds
depends on the scale factor according to
\begin{equation}
G_{N}(R)={\kappa_{5}^{4}|\lambda_{0}|\over 48\pi}R^{-{k^2\over 3}}.
\label{enc}
\end{equation}

In the case $w=-1$ and $k^2\neq 0$, the Friedmann equation is retrieved directly from Eq. (\ref{mpfe}), with the result
\begin{equation}
\left({\dot{R}\over R}\right)^2={1\over 36}(\kappa_{5}^{4}\lambda_{0}^2-36\xi^2)R^{-{2\over 3}k^2}\equiv 
{8\pi\over 3}G_{N}(R)\rho_{Q}(R),
\label{fewmo}
\end{equation}
so the dependence on the matter energy density disappears and 
the dynamics of the universe is controlled only by the induced ``quintessence'' energy density. When $w=-1$ and 
$k=0$, there is an explicit dependence on $\rho_{0}$ and the corresponding Friedmann equation is given
by (\ref{fewmo}) with the replacement $\lambda_{0}\rightarrow \lambda_{0}+\rho_{0}$.

In order to study the solutions of Eq. (\ref{fe}), it is convenient to rewrite it
as the equation of motion of a particle moving with zero total energy in a potential,
\begin{equation}
\dot{R}^2+V(R)=0,
\label{eom}
\end{equation}
where 
\begin{equation}
V(R)=\xi^2 R^{-{2\over 3}(k^2-3)}-{\kappa_{5}^4\over 36}\left[|\lambda_{0}| R^{-{1\over 3}(k^2-3)}
+\rho_{0} R^{-(m-1)}\right]^2.
\label{pot}
\end{equation}
Inflationary solutions on the brane are those with $\dot{R}>0$, $\ddot{R}>0$. Using 
the equations of motion (\ref{eom}), we find that the conditions for inflation reduce to
\begin{equation}
V'(R)<0,
\end{equation}
where the prime denotes differentiation with respect to $R$. 

By looking at Eq. (\ref{pot}), we see that if $m>1$, the brane energy density tends
to stop inflation, whereas the bulk and brane vacuum energies compete with each other. 
When $k^2<3$, the brane vacuum energy contributes to accelerate the expansion, whereas
the bulk cosmological constant tends to decelerate it. If $k^2>3$ the situation is reversed.  
In the following we will study more explicitly 
the main features of the cosmological solutions obtained in the cases when the
matter on the brane is dominated by either vacuum energy ($w=-1$) or a gas of ultrarelativistic
particles $(w={1\over 3})$. 

\subsection{Vacuum branes ($w=-1$)}

Let us first consider the case in which the matter on the brane contributes to the 
energy-momentum tensor as a vacuum energy with the equation of state $p=-\rho$. When $k\neq 0$, we
find from Eq. (\ref{fewmo}) that all dependence on $\rho_{0}$ disappears from the Friedmann 
equation and the potential $V(R)$ simplifies to
\begin{equation}
V(R)=\left(\xi^2-{\kappa_{5}^4\over 36}\lambda_{0}^2\right) R^{-{2\over 3}(k^2-3)}.
\label{pott}
\end{equation}
In order to have solutions to Eq. (\ref{eom}) we must demand 
$\kappa_{5}^4\lambda_{0}^2\geq 36\xi^2$. When the
inequality is saturated ($\kappa_{5}^4\lambda_{0}^2=36\xi^2$) the brane remains at a fixed
position $R(\tau)={\rm constant}$ and we retrieve Minkowski space-time on the brane for any value of 
$k$. The brane tension is given by
\begin{equation}
\lambda(\phi)=\pm {6\xi\over \kappa_{5}^4}\,e^{-{k\over 3}\phi}-\rho_{0}e^{4b\phi}
\end{equation}
and the energy-momentum tensor on the brane is $\tau_{\mu\nu}=-\rho_{0}e^{4b\phi}\eta_{\mu\nu}$ (cf. \cite{fkvmbr}).
Actually, these are the self-tuning solutions of Ref. \cite{st} which preserve four-dimensional Poincar\'e invariance
for any value of the cosmological constant in the bulk.

If, on the other hand, $\kappa_{5}^4\lambda_{0}^{2}>36\xi^2$ we find that $V'(R)<0$ for $k^2<3$ and 
hence inflation takes place. By solving explicitly the equations of motion (\ref{eom}) 
with the potential (\ref{pott}), we find that when $k^2\neq 0$, the scale factor scales
with the proper time as
\begin{equation}
R(\tau) \sim \tau^{3\over k^2}.
\end{equation}
Thus, for $0<k^2<3$, the four-dimensional universe undergoes power-law inflation. 

On the other hand, if $k=0$, the brane tension is constant $\lambda(\phi)=\lambda_{0}$ and
solutions to  Eq. (\ref{eom}) will exist provided 
$\kappa_{5}^4(\lambda_{0}+\rho_{0})^2\geq 36\xi^2$. 
When this inequality is saturated, we find the static Randall-Sundrum model \cite{rs}. 
Otherwise, the four-dimensional metric is not static and we have de Sitter inflation,
\begin{equation}
R(\tau) \sim \exp\left(\tau\sqrt{{1\over 3}\Lambda_{\rm eff}}\right),
\end{equation}
i.e., in this case the expansion is driven by an effective four-dimensional
 cosmological constant  $\Lambda_{\rm eff}=
{\kappa_{5}^4\over 12}{(\lambda_{0}+\rho_{0})^{2}-3\xi^2}>0$.

\subsection{Radiation branes $(w={1\over 3})$}

The second physically interesting case is the one in which the matter on the branes is 
a perfect fluid with equation of state $p={1\over 3}\rho$. The first interesting thing to notice in this case
is that $m=4$ and all the equations become independent of the parameter $b$ measuring the coupling between 
the dilaton and the matter on the brane. This is a consequence of the classical conformal 
invariance of a radiation perfect fluid. Now the potential $V(R)$ is given by
\begin{equation}
V(R)=\xi^2\,R^{-{2\over 3}(k^2-3)}-{\kappa_{5}^4\over 36}\left[|\lambda_{0}|R^{-{1\over 3}(k^2-3)}+\rho_{0}
R^{-3}\right]^2
\end{equation}
with $\lambda_{0}>0$ for $k^2<12$ and $\lambda_{0}<0$ if $k^2>12$.

In order to extract the qualitative behavior of the scale factor, we can study the asymptotic limits for large
and small scale factors. When $k^2<12$, we find that
\begin{equation}
V(R)\sim \left\{
\begin{array}{lll}
\left(\xi^2-{\kappa_{5}^4\over 36}\lambda_{0}^2\right)R^{-{2\over 3}(k^2-3)}, & & R \gg 
\left({\rho_{0}\over |\lambda_{0}|}\right)^{3\over 12-k^2}, \\
 & &  \\
-{\kappa_{5}^4\over 36}\rho_{0}^2R^{-6}, & & R\rightarrow 0. 
\end{array}
\right.
\end{equation}
``Large'' universes will result only if we demand
$\kappa_{5}^4\lambda_{0}^2>36\xi^2$. In this case, we observe that for large values of $R$ the evolution of the scale
factor is identical to the one found for vacuum branes. 
Thus, for $k=0$, the universe 
undergoes de Sitter inflation while for $0<k^2<3$ inflation occurs with the scale factor scaling 
as a power of the
proper time. When $k^2=3$, we find coasting behavior of the scale factor
and for $3<k^2<12$ the expansion of
the four-dimensional universe is decelerated. 

On the other hand, when $R\rightarrow 0$ the evolution is always decelerated 
with $R(\tau)\sim \tau^{1\over 4}$ for 
all $0<k^2<12$. This is the expected evolution of the scale factor of a radiation universe whose dynamics is 
dominated by the $\rho^2$ term in Eq. (\ref{fe}). It is interesting to note that the evolution of the 
universe in this regime is completely 
dominated by the matter energy density, while for large $R$ the expansion rate is determined by a competition
between the bulk cosmological constant (encoded in $\xi$) and the brane vacuum energy as measured by $\lambda_{0}$.

Altogether, we find that for a radiation filled braneworld in a bulk dilaton potential 
with $k^2<12$, the cosmological evolution in four dimensions begins with a $\rho^2$-dominated
phase. After a period of decelerated expansion resulting in a decrease of the density,
the universe will enter a regime in which expansion is driven by the term proportional to $\rho$
in the Friedmann equation. This ``standard'' expansion phase is terminated at some time
in which the matter energy density  decreases below the contribution of the 
energy density from the quintessence term. At this moment, the universe will start its
accelerated expansion. To illustrate the point, in Fig. 1 we have plotted the potential 
$V(R)$ and the scale factor for a particular model in this class
of solutions. In Section 5 we will study this type of brane universe as quintessence
models for the observed acceleration of the universe.

\begin{figure}[h]
\centerline{\epsfxsize=2.5in\epsfbox{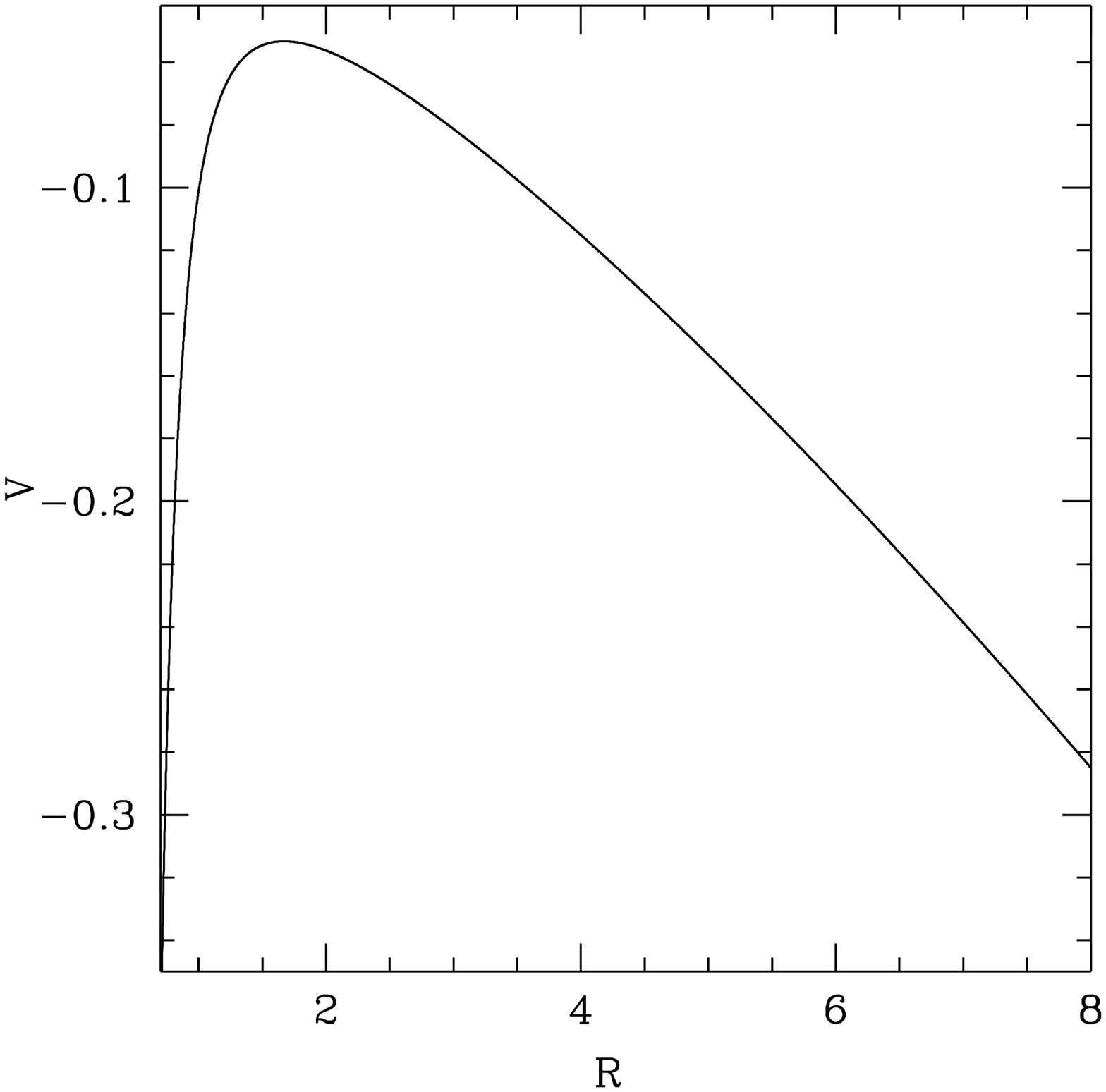} 
\hskip 0.5cm \epsfxsize=2.5in\epsfbox{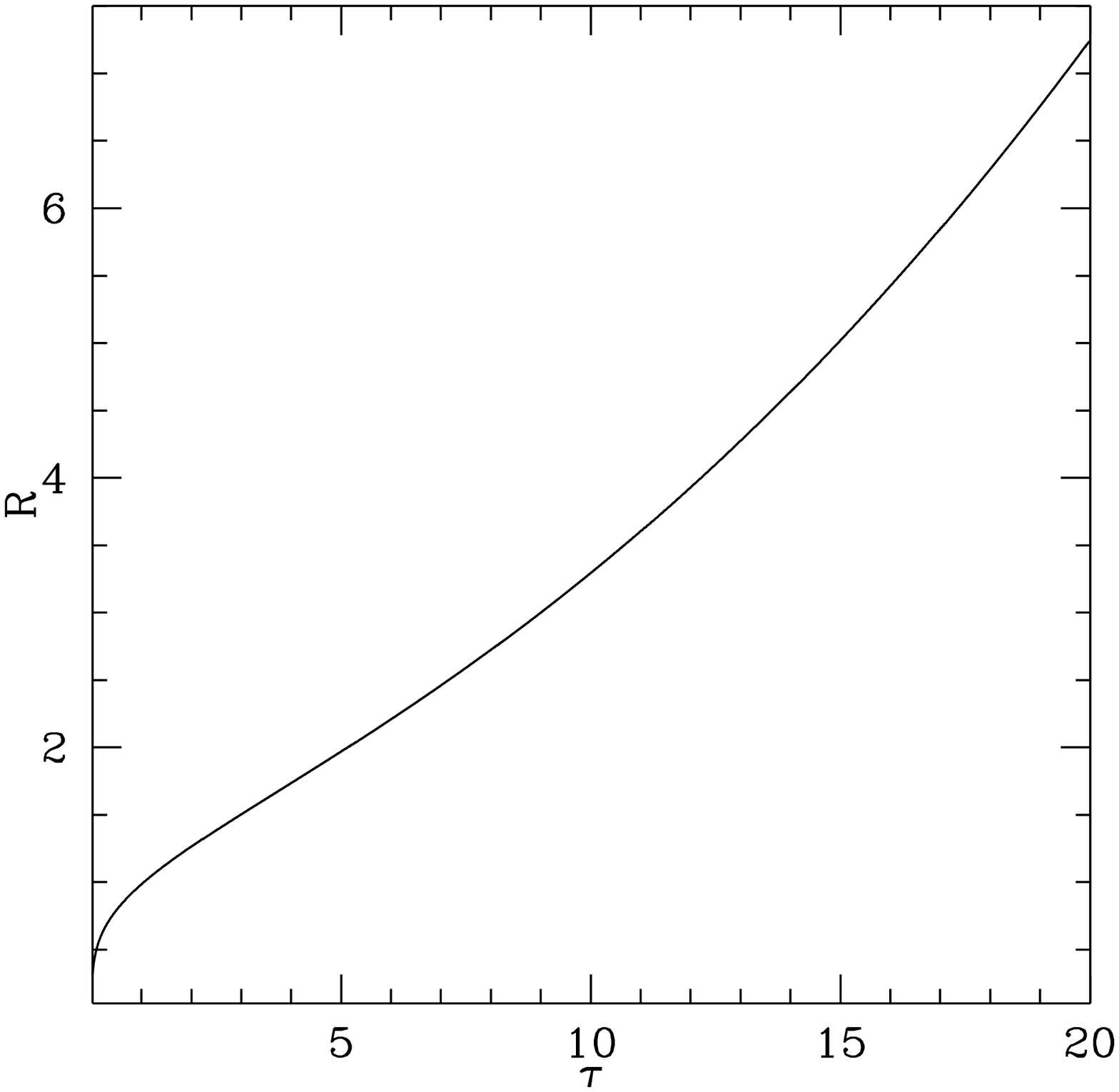}}
\caption{Potential $V(R)$ (left) and numerical solution to the scale factor $R(\tau)$ (right)
for a model with $k=1$, $\xi=0.1$, $\lambda_{0}=1$ and $\rho_{0}=1$. The five-dimensional 
Planck scale is taken $\kappa_{5}=1$.}
\end{figure}

Incidentally, when $k^2<12$ and $\kappa_{5}^4\lambda_{0}^2>36\xi^2$ ($\lambda_{0}>0$) the universe starts at a big-bang
singularity at $\tau=0$ and recollapses after reaching a maximum size of  
$R_{\rm max}= [\kappa_{5}^4\rho_{0}/(6\xi+
\kappa_{5}^4|\lambda_{0}|)]^{3\over 12-k^2}$.

Let us analyze the case when $k^2>12$ which corresponds to a  positive bulk cosmological constant. Now
the asymptotic behavior of the potential is
\begin{equation}
V(R)\sim \left\{
\begin{array}{lll}
-{\kappa_{5}^4\over 36}\rho_{0}^2\,R^{-6}, & & R \gg \left({\rho_{0}\over |\lambda_{0}|}\right)^{3\over12- k^2}, \\
 & &  \\
\left(\xi^2-{\kappa_{5}^4\over 36}\lambda_{0}^2\right)R^{-{2\over 3}(k^2-3)}, & & 
R\rightarrow 0
\end{array}
\right.
\end{equation}
and the situation is reversed with respect to the case analyzed before. 
For large values of $R$, the potential is always negative with $V'(R)>0$ and the decelerated evolution
is controlled by the matter energy density through the $\rho^2$ term in the Friedmann equation (\ref{fe}). 
If again we demand $\kappa_{5}^4\lambda^{2}_{0}<36\xi^2$ ($\lambda_{0}<0$), we find that the universe 
does not have a big-bang singularity, but it starts at a finite size to undergo an inflationary expansion
in which the scale factor grows as a power law. This regime is followed by a ``graceful exit'' phase where
the universe enters a decelerated expansion which asymptotically leads to the $\rho^2$ term dominance. 
An example of this kind of solutions is presented in Fig. 2. In the left panel we have plotted the potential $V(R)$
and on the right one a numerical solution of Eq. (\ref{eom}) is shown. The total amount of inflation is just determined
by the ratio between the size of the universe at the end of inflation $V'(R_{f})=0$ and the original size
of the universe $V(R_{i})=0$.

\begin{figure}[h]
\centerline{\epsfxsize=2.5in\epsfbox{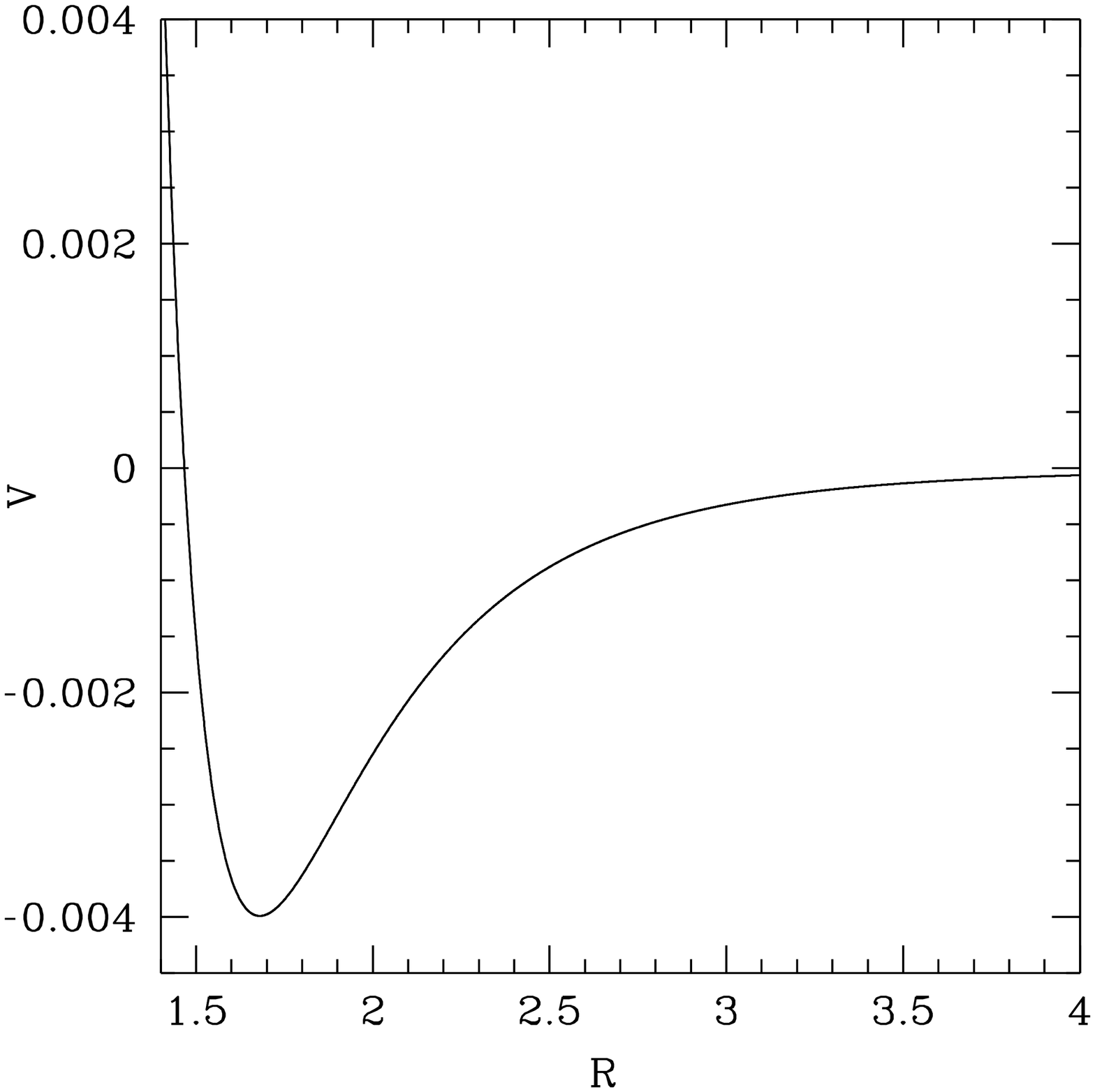} 
\hskip 0.5cm \epsfxsize=2.5in\epsfbox{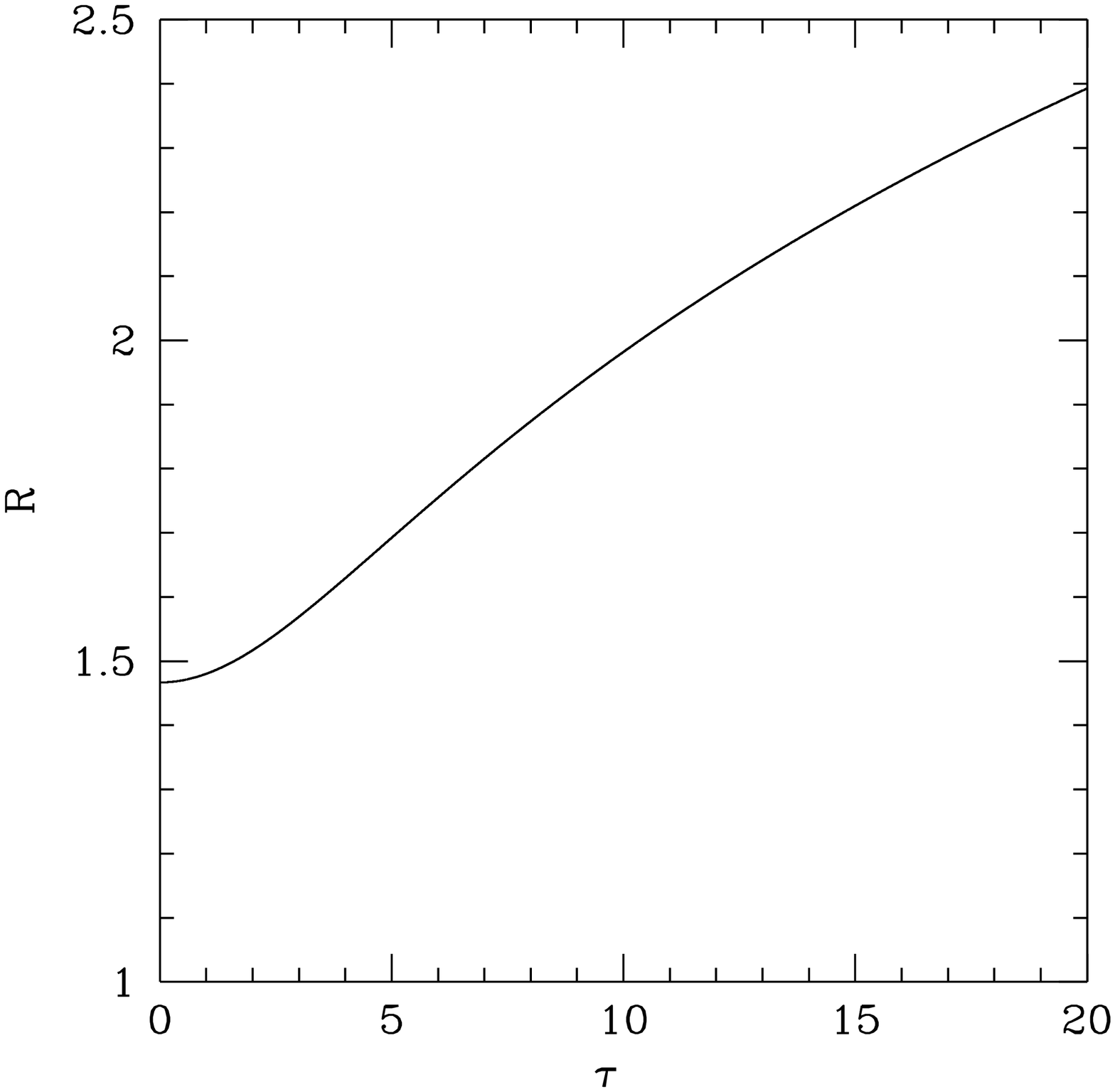}}
\caption{Profile of the potential $V(R)$ (left) and numerical 
solution of the scale factor $R(\tau)$ (right) 
for $k=4$, $\xi=1$, $\lambda_{0}=-1$ and $\rho_{0}=3$. 
We have also taken $\kappa_{5}=1$.}
\end{figure}

If instead we have $\kappa_{5}^4\lambda_{0}^2>36\xi^2$, the universe starts at $R=0$ and experiences a decelerated evolution to
reach asymptotically the $\rho^{2}$-dominated regime. This difference in the behavior when $R\rightarrow 0$ can
be understood by noticing that in the previous case the dynamics of the universe at small values of $R$ is
dominated by a large negative vacuum energy that prevents the formation of the singularity.

Perhaps the most surprising feature of the solutions with $k^2>12$ is the fact that the late-time (large $R$) 
regime is dominated by the $\rho^2$ term in (\ref{fe}). This might seem counter-intuitive, since it is usually argued
that this term should only be important in the very early universe where the energy density of the universe
is very large. The key element to understand the large-$R$ behavior of these solutions is to realize that
in our model the four-dimensional gravitational coupling evolves with time as $R^{-{k^2\over 3}}$,
whereas the $\rho^2$ term couples directly to the five-dimensional Planck scale, which is time independent. 
Thus if $k$ is large enough, both the ``quintessence'' term and the term linear in the energy density become subleading
with respect to the $\rho^2$ contribution when the scale factor grows. Therefore, 
the cosmological evolution is asymptotically dominated by the bulk term in the Friedmann equation.

Before finishing with the analysis of radiation branes, we study,
for the sake of completeness, the marginal case $k^2=12$ where the bulk
dilaton potential vanishes. As explained at the end of Section 2, when $k^2=3m\equiv 12$ the brane tension is given by 
Eq. (\ref{cp}), which contains a logarithmic dependence on $R$. Then $V(R)$ is given by
\begin{equation}
V(R)_{k^2=12}=\left[\xi^{2}-{\kappa_{5}^4\over 36}\left(\lambda_{0}+\rho_0
-4\rho_{0}\log R\right)^2\right]\,R^{-6}.
\end{equation}
This implies that for both large and small $R$, the dynamics of the scale factor is controlled by the
logarithmic term in the potential. The profile of the potential presents then two allowed
regions for a zero-energy particle, for large and small $R$, separated by a  barrier.

\section{Cosmologies on the brane in bulks with broken $\mathbb{Z}_2$-reflection symmetry}
\setcounter{equation}{0}

As we discussed in Section 2, braneworlds can be also embedded in a bulk metric of the form (\ref{lefd}) without
$\mathbb{Z}_{2}$-reflection symmetry by considering different values of the energy
scale $\xi$ at the two sides of the brane location. In this case, the evolution equations for the
scale factor are obtained by solving Eq. (\ref{ehh}) for the Hubble parameter with the result
\begin{equation}
\left({\dot{R}\over R}\right)^2=-{\xi_{+}^2+\xi_{-}^2\over 2}
R^{-{2\over 3}k^2}+
{\kappa_{5}^{4}\over 36}(\lambda+\rho)^2
+{9\over 4\kappa_{5}^{4}}\left(\xi_{+}^2-\xi_{-}^2\right)^2
(\lambda+\rho)^{-2}R^{-{4\over 3}k^2}
.
\label{ee2}
\end{equation}
Comparing with Eq. (\ref{pfe}), we notice the presence of an extra term which scales with a negative power of
the brane energy density (cf. \cite{davis}). The expressions for $\rho(R)$ and $\lambda(\phi)$ are the same as for the symmetric
case. Proceeding as in the previous section, we reduce the problem of integrating the scale factor to 
finding the trajectory of a classical nonrelativistic particle with zero energy in the potential,
\begin{eqnarray}
V(R)&=&{\xi_{+}^2+\xi_{-}^2\over 2}R^{-{2\over 3}(k^2-3)}
-{\kappa_{5}^4\over 36}\left[|\lambda_{0}| R^{-{1\over 3}(k^2-3)}
+\rho_{0} R^{-(m-1)}\right]^2 \nonumber \\
&-&{9\over 4\kappa_{5}^4}\left(\xi_{+}^2-\xi_{-}^2\right)^2
R^{-{4\over 3}(k^2-3)}
\left[|\lambda_{0}| R^{-{1\over 3}(k^2-3)}
+\rho_{0} R^{-(m-1)}\right]^{-2},
\label{eee2}
\end{eqnarray}
where, as above, $\rho_{0}R^{-m}$ represents the rescaled effective energy density on the
brane. We notice that the last extra term in the potential
associated with the breaking of reflection symmetry is always negative and therefore its contribution tends to 
increase the value of $H^{2}\equiv\left(\frac{\dot{R}}{R}\right)^2$. Moreover, it can be easily checked that this new 
term is bounded by the second one
on the right-hand side of Eq. (\ref{eee2}) for all values of $R$ provided 
$|\lambda_{0}|\geq \frac{3}{\kappa_5^2}\sqrt{|\xi_{+}^2-\xi_{-}^2|}$. 
As a consequence, when this bound is
satisfied the dynamics of the brane universe is never dominated by the extra term in Eq. (\ref{ee2}) and the
cosmological evolution is qualitatively similar to the one found in the previous section.

In order to show the qualitative differences with respect 
to the symmetric case studied in the previous section, 
we have to evaluate the relevance of the extra term in Eq. (\ref{ee2}). Here we will focus on the solutions with 
phenomenological applications, therefore, for reasons that will be explained in Section 5, we 
restrict our attention to  cases 
with $k^2<3m$. If $|\lambda_{0}|\geq \frac{3}{\kappa_5^2}
\sqrt{|\xi_{+}^2-\xi_{-}^2|}$, the extra term is negligible at small $R$ 
compared with the dominating $\rho^2$ term. 
In general, however, accelerated expansion takes place for 
large values of $R$. In this latter case,
the effective four-dimensional
vacuum energy depends on the scale of $\mathbb{Z}_{2}$-symmetry breaking,
$(\xi_{+}^2-\xi_{-}^2)^2$,
\begin{equation}
V(R)\sim\left\{ 
\begin{array}{lll}
\left[{\xi_{+}^2+\xi_{-}^2\over 2}-{\kappa_{5}^4\over 36}\lambda_{0}^2-
{9(\xi_{+}^2-\xi_{-}^2)^2\over 4\kappa_{5}^4\lambda_{0}^2}\right]R^{-{2\over 3}
(k^2-3)}, & & R\gg \left({\rho_{0}\over |\lambda_{0}|}\right)^{3\over 3m-k^2}, \\
  & &  \\
-{\kappa_{5}^2\over 36}\rho_{0}^2R^{-2(m-1)}, & & 
R\rightarrow 0.
\end{array}
\right.
\end{equation}

Taking $\kappa_{5}^4\lambda_{0}^2[18(\xi_{+}^2+\xi_{-}^2)-
\kappa_{5}^4\lambda_{0}^2]<81(\xi_{+}^2-\xi_{-}^2)^2$ it is easy 
to check that  
the universe recollapses within  finite proper time without passing 
through a phase of accelerated expansion. Therefore, these solutions are not
very interesting from a phenomenological point of view. On the other hand, if 
$\kappa_{5}^4\lambda_{0}^2[18(\xi_{+}^2+\xi_{-}^2)-
\kappa_{5}^4\lambda_{0}^2]>81(\xi_{+}^2-\xi_{-}^2)^2$, the universe 
starts again in a $\rho^2$-dominated phase that, when $\rho$ decreases, will be followed by a
standard expansion dominated by the term linear in $\rho$ in Eq. (\ref{ee2}). Eventually, however, the (positive) 
quintessence energy density will dominate and the universe enters a regime of 
accelerated expansion. 

Actually, when $R$ is large enough and higher powers of
$\rho$ can be neglected 
[which includes the limit $R\gg (\rho_{0}/|\lambda_{0}|)^{3\over 3m-k^2}$] 
the dynamics of the universe  
can  be derived from the approximate Friedmann equation,
\begin{equation}
\left({\dot{R}\over R}\right)^2\approx {8\pi\over 3}G_{N}(R)\rho_{Q}(R)+{8\pi\over 3}G_{N}(R)\rho(R),
\label{afe}
\end{equation}
where the new Newton constant is given by
\begin{equation}
G_{N}(R)\approx {\kappa_{5}^4|\lambda_{0}|\over 48\pi}
\left[1-{81\over 2\kappa_{5}^8\lambda_{0}^4}(\xi_{+}^2-\xi_{-}^2)^2\right]R^{-\frac{k^2}{3}}
\end{equation}
and the quintessence component of the energy density is 
\begin{equation}
\rho_{Q}(R)\approx {1\over 2\kappa_{5}^4|\lambda_{0}|}
\left[\kappa_{5}^4\lambda_{0}^2-18(\xi_{+}^2+\xi_{-}^2)+
{162\over \kappa_{5}^4\lambda_{0}^2}
\left(1-9{\xi_{+}^2+\xi_{-}^2\over \kappa_{5}^4\lambda_{0}^2}\right)
(\xi_{+}^2-\xi_{-}^2)^2
\right]R^{-{k^2\over 3}}.
\end{equation}
For consistency, only terms linear in ${1\over \kappa_{5}^{4}\lambda_{0}^2}(\xi_{+}-\xi_{-})^2$ are retained 
upon substitution in Eq. (\ref{afe}).
We see that both  Newton's constant and the quintessence energy density 
reduce to  expressions (\ref{quint}) and (\ref{enc})
for $\xi_{+}=\xi_{-}$.

Let us consider now the case in which the difference between the cosmological 
constant at the two sides of the 
brane is ``large'', i.e. $\kappa_{5}^2|\lambda_{0}|< 3\sqrt{|\xi_{+}^2-\xi_{-}^2|}$. 
Now the last term in Eq. (\ref{ee2}) will dominate
for a certain range of values of $R$. Thus, the $\rho^2$-dominated epoch at 
small $R$ will be followed 
by a regime where the expansion is driven by the term proportional to $(\lambda+\rho)^{-2}$. This will 
happen for the following range of values of the scale factor:
\begin{equation}
\left({\kappa_{5}^2\rho_{0}\over 3\sqrt{|\xi_{+}^2-\xi_{-}^2|}}\right)^{3\over 3m-k^2}
\lesssim R \lesssim \left({\rho_{0}\over 
|\lambda_{0}|}\right)^{3\over 3m-k^2}.
\end{equation}
For larger values of $R$, the universe will enter a quintessence-dominated regime.

To summarize, we have seen that when the difference between the values of $\xi$ at both
sides of the brane is small (in the sense explained above), the cosmological dynamics
of the universe does not show a radically different
behavior with respect to the symmetric models. Actually, for large values of
$R$ the effect of the breaking of the reflection symmetry shows in a correction of 
both Newton's constant and $\rho_{Q}$, although their scalings with
the scale factor are not affected. On the other hand, for larger values of $|\xi_{+}-\xi_{-}|$,
the main difference is the existence of an intermediate phase interpolating between the 
$\rho^2$ regime and the ``standard" expansion dominated by the linear term in the Friedmann
equation, where the dynamics of the universe is controlled by the last term in Eq. (\ref{eee2}).
At large values of $R$, however, the dynamics can still be described by the approximate
Friedmann equation (\ref{afe}).

\section{Phenomenology of quintessential brane cosmologies}
\setcounter{equation}{0}

After having investigated their mathematical properties, 
in the following we will study  phenomenological aspects 
of the quintessential brane cosmological models presented in 
Section 3, where the acceleration of the universe at late times is induced by the rolling of the braneworld in a static bulk 
geometry. We will focus our attention on the $\mathbb{Z}_{2}$-reflection symmetric models. 

We saw that whenever
$\kappa_{5}^4\lambda_{0}^2> 36\xi^2$ and $k^2<3m$ (with $m>0$), solutions 
describing a flat
Friedmann-Robertson-Walker universe can be found
which, after going through radiation/matter-dominated
epochs, undergo accelerated expansion at late times.  
For these types of solutions, the energy content of
the very early universe is dominated by radiation and the expansion is 
controlled by the $\rho^2$ term in Eq. (\ref{fe}).
When the density is low enough, a phase of ``standard'' expansion starts where the
dynamics of the universe is governed by the term linear in $\rho$
in the Friedmann equation. It is during this
phase that nucleosynthesis has to take place, as well as the decoupling of matter from radiation. Finally, some time 
after the universe enters in a matter dominated epoch, the quintessence term starts dominating the dynamics and the 
universe begins its accelerated expansion. This acceleration is driven by the 
bulk scalar field through the term $\rho_{Q}$ in the Friedmann equation, which, as we will see below, 
plays the role of effective quintessence on the brane. In addition to this, the effective Newton constant in our models, 
Eq. (\ref{enc}), evolves with time through its dependence on the scale factor.

We normalize the scale factor on the brane so it takes the value $R_{0}=1$ at the
present time
and we use the temperature of the radiation to label the different
epochs in the evolution of the universe,
\begin{equation}
T(R)={2.73\,{\rm K}\over R}.
\end{equation}
From our results in Sec. 2.3 we saw that the energy density on the brane scales with the 
temperature as 
\begin{equation}
\rho(T)=\rho_{0}\left({T\over 2.73\,{\rm K}}\right)^{3(1+w_{\rm eff})},
\end{equation}
where $w_{\rm eff}$ is give by Eq. (\ref{meos}). As discussed above, for the particular
case of radiation $w_{\rm eff}=w={1\over 3}$, and we recover Boltzmann's law, $\rho(T)\simeq
T^{4}$. However, when $w=0$ the effective barotropic index is nonvanishing. Nonetheless, in
the following we will denominate ``matter'' to any perfect fluid with $w_{\rm eff}=0$, whose energy
therefore scales with the temperature as $\rho(T)\simeq T^{3}$, as seen by a brane observer.
Assuming by naturality that $b\simeq k$, we find that $w_{\rm eff}-w\simeq k^2$. As we will see
later, this number is of order $\lesssim 10^{-2}$ so the difference between the ``microscopic'' 
barotropic index and the effective one (which governs the cosmological 
evolution) is very small. 
  
To study the evolution of the universe in more quantitative terms, it is useful to introduce
three parameters, $\Theta_{Q}(T)$, $\Theta_{\rho}(T)$, and $\Theta_{\rho^2}(T)$, which measure the contribution to the 
expansion of the three terms on the right hand side of the Friedmann equation (\ref{fe}),
normalized to the Hubble parameter,
\begin{eqnarray}
\Theta_{Q}(T)&=& {8\pi\over 3H^2}G_{N}(T)\rho_{Q}(T),\nonumber \\
\Theta_{\rho}(T)&=& {8\pi\over 3H^2}G_{N}(T)\rho(T), \label{opar}\\
\Theta_{\rho^2}(T)&=& {\kappa_{5}^{4}\rho(T)^2\over 36 H^2},
\nonumber 
\end{eqnarray}
where $\rho_{Q}(T)$ is given by Eq.(\ref{quint}). In this way the Friedmann equation is written as
\begin{equation}
1=\Theta_{Q}(T)+\Theta_{\rho}(T)+\Theta_{\rho^2}(T).
\end{equation}

The crossover between the $\rho^2$-dominated epoch and the standard regime will take place at a temperature
$T_{\rm c}$ at which $\Theta_{\rho^2}(T_{c})\simeq \Theta_{\rho}(T_{\rm c})$. By using the definitions 
(\ref{opar}) we find
\begin{equation}
{T_{\rm c}\over 2.73\,{\rm K}}\simeq \left({2|\lambda_{0}|\over \rho_{0}}\right)^{3\over 12-k^2}.
\label{Tc}
\end{equation}
On the other hand the standard expansion of the universe will be terminated at a temperature $T_{Q}<T_{\rm c}$ in which 
$\Theta_{\rho}(T_{Q})\simeq \Theta_{Q}(T_{Q})$. Since this has to occur after the universe has entered a matter-dominated 
era, we find
\begin{equation}
{T_{Q}\over 2.73\,{\rm K}} \simeq \left({
\kappa_{5}^4\lambda_{0}^2-36\xi^2\over 2\kappa_{5}^{4}|\lambda_{0}|\rho_{0}}\right)^{3\over 9-k^2}.
\label{TQ}
\end{equation}
As a consequence, the period of standard expansion can be very large
if the net cosmological constant on the 
brane, $\kappa_{5}^{4}\lambda_{0}^2- 36\xi^2$, is small enough.

In order to illustrate the discussion above, in Fig. 3 
the evolution of the Hubble parameter is shown for some ``realistic'' values of
the parameters in a cold dark matter model.
We can see clearly several regimes. At high temperatures ($T\gtrsim 10^{12}\,{\rm K}$),
the expansion is dominated by the $\rho^2$ term in the Friedmann equation. This is followed
by a long period of standard expansion for $10^{12}\,{\rm K}\lesssim T \lesssim 10\,{\rm K}$.
In this regime, radiation ($m=4$) dominates the expansion for temperatures higher 
than $T_{\rm eq}\simeq
10^{4}\,{\rm K}$, whereas for lower temperatures the energy density of the universe is
dominated by matter ($m=3$). Finally, when $T\lesssim 10\,{\rm K}$, the quintessence component
takes over and the universe starts its accelerated expansion.

\begin{figure}[h]
\centerline{\epsfxsize=3.5in\epsfbox{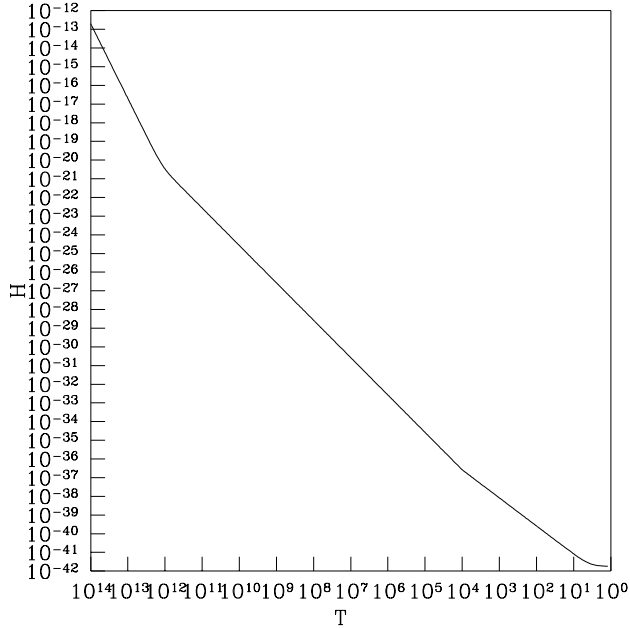}}
\caption{Evolution of the Hubble parameter (GeV) vs. temperature (K) for a realistic quintessence model.
The values of the parameters are:
$\rho_{0}=0.43 \rho_{Q,0} = 2.43 \times 10^{-47}\,{\rm GeV}^4$, 
$G_{N,0}=6.71 \times 10^{-39}\,{\rm GeV}^{-2}$, $T_{\rm eq}=10^{4}\,$K,
$k^2=0.01$ and $\kappa_{5}=10^{-8}\,{\rm GeV}^{-{3\over 2}}$.}
\end{figure}

\subsection{Bounds from big-bang nucleosynthesis}

The abundances with which light elements are synthesized in the early universe are extremely sensitive to changes in the 
expansion rate of the universe and/or in the fundamental constants \cite{peacock,prep}. Thus, primordial nucleosynthesis provides a 
good way to bound the values of the constants and scales that characterize a given cosmological model. 
In the case at hand, the effective gravitational coupling varies with time and from Eq. (\ref{enc}) we can see that this 
change is determined solely by $k$, namely
\begin{equation}
{\dot{G}_{N}\over G_{N}} =-{k^2\over 3} H.
\end{equation}
In order to explain the primordial abundances of $^{4}$He and other light elements, however, Newton's constant at the time of
nucleosynthesis cannot be very different from the present value. Actually, its time variation has to satisfy  \cite{bbn}
\begin{equation}
\left({\dot{G}_{N}\over G_{N}}\right)_{0} \lesssim \pm 0.01\,h^{-1}\,H_{0}.
\label{b1}
\end{equation}
Therefore, we find that the parameter $k$ that determines the slope of the potential of the bulk dilaton has to be rather small,
\begin{equation}
k^{2} \lesssim 0.03\,h^{-1},
\label{bound1}
\end{equation}
which, from Eq. (\ref{wQ}), implies $w_{Q}+1\lesssim 0.003\,h^{-1}$ (cf. \cite{corcop}). 

A second condition to be met by a cosmological model 
in order to account for the primordial abundances of
the light elements is that the energy density of the universe at the time of nucleosynthesis has to be dominated by
radiation. In particular, the dark energy contribution to the density parameter 
satisfies the following bound \cite{bhm} (see also \cite{ferjoy} and references therein):
\begin{equation}
\Omega(T_{\rm BBN})_{\rm dark} < 0.045 \hspace*{1cm} \hbox{(at 2$\sigma$ confidence level)},
\label{b2}
\end{equation}
where $T_{\rm BBN}\simeq 10^{10}{\rm K}$. In our model, the dark energy component corresponds to the contribution of $\rho_{Q}$. Thus, we define
\begin{equation}
\Omega_{Q}(T)={\rho_{Q}(T)\over \rho_{Q}(T)+\rho(T)},
\label{omega}
\end{equation}
where $\rho(T)$ is the matter/radiation energy density. 

From Eq. (\ref{omega}), and assuming a cold dark matter scenario, 
we can obtain an expression of $\Omega_{Q}(T_{\rm BBN})$ in terms of quantities at
the present time as
\begin{equation}
\Omega_{Q}(T_{\rm BBN})=\left[1+{1-\Omega_{Q}(T_{0})\over \Omega_{Q}(T_{0})}\left({T_{\rm eq}\over T_{0}}\right)^{3}\left({T_{\rm BBN}
\over T_{\rm eq}}\right)^{4}\left({T_{\rm BBN}\over T_{0}}\right)^{-{k^2\over 3}}\right]^{-1}  ,
\label{omegaBBN}
\end{equation}
where $T_{\rm eq}\simeq 10^{4}\,{\rm K}$ is the temperature at which the density of radiation and matter are equal, and $T_{0}= 2.73 \,{\rm
K}$ is the temperature of the CMB at the present time. Since the measurement of CMB anisotropies and the 
supernova data indicate that $\Omega_{Q}(T_{0})\simeq 0.7$,
we find that for  models satisfying the bound (\ref{bound1}) the contribution of the dark energy component at the time of 
nucleosynthesis is 
\begin{equation}
\Omega_{Q}(T_{\rm BBN})\simeq 10^{-35},
\end{equation}
which falls well inside the nucleosynthesis bound (\ref{b2}).

\subsection{Bounds from observations}

We have seen that by satisfying the bound (\ref{bound1}), our quintessential brane cosmologies comply with the 
constraints imposed by big-bang nucleosynthesis. It is interesting to notice, however, that none of 
the two bounds considered in the previous section involves any other parameter of the model
but the numerical constant $k$, or equivalently the quintessence effective barotropic index
$w_{Q}$. 
In particular we do not get any information as to the size of the dimensionful parameters
$\lambda_{0}$ and $\xi$ or the five-dimensional Planck scale $\kappa_{5}$. 

In order to extract bounds for these scales, we have to compare the values of $\rho_{Q}$ and 
the present value of the four-dimensional Newton constant 
with the observed values. From Eq. (\ref{enc}) we get the value of the Newton's constant 
by setting $R_{0}=1$. Using $G_{N,0}= 6.71\times 10^{-39}\,{\rm GeV}^{-2}$, we find
\begin{equation}
\kappa_{5}^{4}|\lambda_{0}|=6\kappa_{4}^2\simeq 10^{-36}\,{\rm GeV}^{-2}.
\label{est1}
\end{equation}

The second observational result we can use to estimate the scales of the model is the value of the 
dark energy component of the density parameter, 
$\Omega(T_{0})_{\rm dark}\simeq 0.7$. From Eqs. (\ref{omega}) and (\ref{quint}), we find that 
\begin{equation}
\kappa_{5}^4\lambda_{0}^2-36\xi^2 \simeq 4.67 \kappa_{5}^4|\lambda_{0}|\rho_{0},
\label{kappa}
\end{equation}
where $\rho_{0}$ is the matter density of the universe at the present time. Using 
Eq. (\ref{est1}) and the value
$\rho_{0}\simeq 2.43\times 10^{-47}h^2\,{\rm GeV}^4$ \cite{peacock} we arrive at
\begin{equation}
\kappa_{5}^4\lambda_{0}^2-36\xi^2 \simeq 10^{-82}\,{\rm GeV}^2.
\label{est2}
\end{equation}
Incidentally, using Eqs. (\ref{kappa}) and (\ref{TQ}) we find that $T_{Q}\simeq 3.63\,{\rm K}$.

Equations (\ref{est1}) and (\ref{est2}) are the only two estimates that we can get just using ``in-brane'' information. In order
to go forward and get some estimate for the five-dimensional Planck scale $\kappa_{5}$ or the brane tension $\lambda_{0}$, we need
some knowledge of the fundamental energy scale in the higher-dimensional theory. The $\rho^2$ term in the Friedmann equation
(\ref{fe}) provides a window to the bulk dynamics, since it couples directly to the five-dimensional Planck 
scale $\kappa_{5}$. Therefore, it would be very interesting to have some
information about the transition between the $\rho^2$-dominated epoch and the standard expansion. Even if we do not have
a concrete value for the temperature at which this transition takes places, we can still 
get a bound for $\lambda_{0}$ by going back to Eq. (\ref{Tc}). 
In order to preserve big-bang nucleosynthesis,  
the $\rho^2$-component in the Friedmann equation has 
to be negligible at $T_{\rm BBN}$ \cite{carkap}. 
Imposing $T_{\rm c}\gg
10^{10}\,{\rm K}$, we are led to
\begin{equation}
|\lambda_{0}|\gg 10^{40}\rho_{0} \simeq 10^{-7} \,{\rm GeV}^{4}.
\end{equation}
Using  Eq. (\ref{est1}) the following bound for the five-dimensional Planck mass
is found, 
\begin{equation}
M_{5} \equiv \kappa_{5}^{-{2\over 3}}\gg 10^{5}\,{\rm GeV},
\label{5dpm}
\end{equation}
and consequently $\xi\gg 10^{-22}\, {\rm GeV}$.
Notice that this bound for the five-dimensional Planck mass 
is stronger than the one 
obtained in the quintessence scenario of Ref. \cite{mizmae}, although it is again weaker than the one obtained from the 
Randall-Sundrum model \cite{rs}.

\section{Conclusions}
\setcounter{equation}{0}
In the present paper, we have studied the bulk solutions constructed in Ref. \cite{fkvmbr} 
in a different physical setup. Instead of being at a fixed 
location, the braneworlds move in a five-dimensional 
dilatonic bulk space time with and without 
$\mathbb{Z}_{2}$-reflection symmetry.
As a result, and contrary to the case analyzed in Ref. \cite{fkvmbr}, 
the cosmological dynamics is not driven by a time-dependent background 
but by the movement of the braneworld in an otherwise static bulk space-time. 
The resulting class of brane cosmologies 
admits, depending on the choice of parameters,
either inflationary behavior at early times and decelerated expansion at
late times or a decelerated expansion at early times which  evolves into
a quintessencelike behavior at late, i.e., present, times.

The realization of quintessence in these models seems to be particularly
interesting since it does not rely on the introduction of an additional
scalar field on the brane. It is in fact driven by the bulk dilaton.
As the brane moves along in the static bulk, it sweeps out sections with different values for
the  bulk dilaton and, from the point of view of an observer living on the brane,  this is interpreted as a
time-dependent cosmological constant. For a suitable
choice of parameters, this evolution is such that the
energy density of this effective quintessence only becomes
important at the present time. However, in addition to the 
induced time-dependent cosmological constant the
bulk dilaton also causes a time-dependent gravitational
coupling.
Using observational constraints on the variation of Newton's
constant and the present value of the cosmological constant
gives bounds on the parameter of the brane cosmologies at hand.
The fact that
$\kappa_{5}^4\lambda_{0}^2-36\xi^2\ll \kappa_{5}^4\lambda_{0}^2,\xi^2$ 
seems to indicate that the nonvanishing value of the four-dimensional vacuum 
energy is the result of a slightly broken symmetry.
Furthermore, we find a lower limit for the five-dimensional
Planck mass of $10^{5} \,{\rm GeV}$. 

Here we are taking a phenomenological approach and therefore we do not
attempt to explain these values for the scales of the model from first
principles. However, it would be very interesting to investigate whether
the hierarchy of scales found here can be obtained in a natural way by 
embedding these five-dimensional models in a string/M-theory setup.

We have shown that for a certain  range of parameters, 
inflation takes place in the early universe. However, this phase is
followed, after 
a graceful exit period, by a decelerated Friedmann-Robertson-Walker
universe in which the expansion is driven by $\rho^2$. This fact
makes this particular set of solutions not very promising as phenomenological
models of the universe \cite{pheno}.

We have also considered brane cosmologies with broken $\mathbb{Z}_{2}$-reflection
symmetry and found that for a slight symmetry breaking the evolution follows qualitatively that
of the symmetric case. In this case, for large values of the scale factor the 
evolution is approximately described by a Friedmann equation where both
Newton's constant and the quintessence energy density receive corrections 
depending on the scale of $\mathbb{Z}_{2}$-symmetry breaking.

One of the interesting properties of the braneworlds 
investigated in \cite{fkvmbr} is the existence of an 
extension of the mechanism of self-tuning \cite{st} in which the matching
conditions at the brane location force the cancellation of all 
vacuum energy terms on the 
right-hand side of the Einstein equations in four dimensions. 
However, from our analysis in Section 3 we find that the 
physical mechanism enforcing this self-tuning is
the condition that the braneworld is at a fixed position in the bulk. 
Indeed, from Eq. (\ref{pfe}) we conclude that the condition $\dot{R}=0$ 
for any $R$ is equivalent to consider a vacuum brane ($w=-1$) 
with zero net cosmological constant in four dimensions, 
$\kappa_{5}^4\lambda_{0}^2=36\xi^2$. From this point of view, at least 
in the case of a static bulk geometry, the self-tuning can be seen 
as resulting from the tuning of the velocity of the brane in the bulk 
to zero\footnote{ This result extends to the cases with 
$\mathbb{Z}_{2}$-reflection symmetry breaking, where again 
a nonmoving brane is only achieved by considering a vacuum brane
with $\kappa_{5}^4\lambda_{0}^2=9(\xi_{+}\pm\xi_{-})^2$.}. 
However, this is not necessarily a fine-tuning, 
since the condition $\dot{R}=0$ can be automatically implemented
if the braneworld is confined to lie on the fixed points of 
an $\mathbb{S}^{1}/\mathbb{Z}_{2}$
orbifold, as it is the case in Ho\v{r}ava-Witten type models \cite{hw}.

\section*{Acknowledgments}

It is a pleasure to thank Alex Feinstein for interesting discussions and 
a most enjoyable collaboration in Ref. \cite{fkvmbr}.
K.E.K. has been supported by the Swiss National Science Foundation and Spanish Science Ministry Grant 
1/CICYT 00172.310-0018-12205/2000. The work of M.A.V.-M. has been
supported by EU Network ``Discrete Random Geometry'' Grant HPRN-CT-1999-00161, 
ESF Network no. 82 on ``Geometry and Disorder'', 
Spanish Science Ministry Grant AEN99-0315, and 
University of the Basque Country Grants UPV 063.310-EB187/98 
and UPV 172.310-G02/99. K.E.K. thanks the Niels Bohr Institute for hospitality.
M.A.V.-M. thankfully acknowledges the hospitality of the Theory Division at CERN 
and the Cosmology Group at the Physics Department of 
the University of Geneva during the 
completion of this work.


\setlength{\baselineskip}{6mm}

\end{document}